\documentclass[%
 reprint,
nofootinbib,
 amsmath,amssymb,
 aps,
]{revtex4-1}

\usepackage{graphicx}% Include figure files
\usepackage{dcolumn}% Align table columns on decimal point
\usepackage{bm}% bold math
\usepackage{graphics}
\usepackage[T1]{fontenc} % if needed
\usepackage{rotating}
\usepackage{lipsum}
\usepackage{graphicx,slashed,hyperref,color,subfig}
\usepackage{amssymb}
\usepackage[normalem]{ulem}
\usepackage[utf8]{inputenc}
\usepackage{ulem} % \sout{ strikeout}
 \hypersetup{
   bookmarks=true,         % show bookmarks bar?
   unicode=true,          % non-Latin characters in Acrobat�s bookmarks
   pdftoolbar=true,        % show Acrobat�s toolbar?
   pdfmenubar=true,        % show Acrobat�s menu?
   pdffitwindow=false,     % window fit to page when opened
   pdfstartview={FitH},    % fits the width of the page to the window
   pdftitle={My title},    % title
   pdfauthor={Author},     % author
   pdfsubject={Subject},   % subject of the document
   pdfcreator={Creator},   % creator of the document
   pdfproducer={Producer}, % producer of the document
   pdfkeywords={keyword1} {key2} {key3}, % list of keywords
   pdfnewwindow=true,      % links in new PDF window
   colorlinks=true,       % false: boxed links; true: colored links
   linkcolor=blue,          % color of internal links (change box color with linkbordercolor)
   citecolor=magenta,        % color of links to bibliography
   filecolor=magenta,      % color of file links
   urlcolor=cyan           % color of external links
 }

\def\lsim{\mathrel{\rlap{\lower4pt\hbox{\hskip1pt$\sim$}}
    \raise1pt\hbox{$<$}}}         %less than or approx. symbol
\def\gsim{\mathrel{\rlap{\lower4pt\hbox{\hskip1pt$\sim$}}
    \raise1pt\hbox{$>$}}}         %greater than or approx. symbol

\begin{document}
%\preprint{APS/123-QED}

\title{ Probing new physics with high-multiplicity events: \\ 
UHE neutrinos at air-shower detector arrays}

\author{Yongsoo Jho}
 \email{jys34@yonsei.ac.kr}
\affiliation{Department of Physics and IPAP, Yonsei University, Seoul 03722, Republic of Korea}

\author{Seong Chan Park}
 \email{sc.park@yonsei.ac.kr}
\affiliation{Department of Physics and IPAP, Yonsei University, Seoul 03722, Republic of Korea}

\date{\today}% It is always \today, today,

\begin{abstract}
Semi-classical processes such as production and decay of electroweak sphaleron in the Standard Model and also microscopic black hole in low scale gravity scenario typically involve large number of particles in final states. These large multiplicities can be distinctively seen in collisions of  Ultra-high-energy (UHE) neutrinos with $E_\nu \gsim 10^9~{\rm GeV}$ and nucleons in the atmosphere of the Earth. Focusing on air-shower detector array experiments including Telescope Array Experiment (TA), Pierre-Auger Observatory (Auger), we propose strategic ways to discover and analyze such events.

%\begin{description}
%\end{description}
\end{abstract}
\pacs{Valid PACS appear here}% PACS, the Physics and Astronomy
\maketitle
%\tableofcontents

%%%%%%%%%%%%%%%%%%%%%%%
\section{Introduction\label{sec:introduction}}
%%%%%%%%%%%%%%%%%%%%%%%

Recent observation of Ultra-High-Energy cosmic rays (UHECRs) extended the domain of the high-energy frontier covering beyond the reach of collider experiments. Many $\mathcal{O}(10)$ EeV cosmic ray events and a few $\mathcal{O}(1)$ PeV neutrino events have been already observed by ground air-shower detector arrays~\cite{Valino:2015zdi,Tinyakov:2018hfg} and other large detectors~\cite{Aartsen:2013bka}. There are widely accepted and also more speculative potential origins of such UHECRs: young magnetized neutron star~\cite{Blasi:2000xm}, active galactic nuclei~(AGN) \cite{Halzen:1997hw}, gamma-ray burst (GRB)~\cite{Waxman:1995dg,Vietri:1995hs}, BL Lac type objects \cite{Gorbunov:2004bs}, dark matter and topological defects \cite{Bhattacharjee:1998qc} but none of them has been confirmed yet. Furthermore, the fluxes, composition and the production mechanisms of UHECRs sensitively depend on the origin. Thus they still need to be clarified~\cite{Aartsen:2015zva,Ishihara:2016pty,Abreu:2013zbq,Aab:2015kma,Abbasi:2018nun,TheTelescopeArray:2018dje}. {From the observation of UHECRs up to $E_{\rm CR} \sim 10^{21}$ eV, we also naturally expect the \textit{guaranteed} existence of Ultra-High-Energy (UHE) neutrinos, produced by the Greisen-Zatsepin-Kuzmin (GZK) mechanism between UHECR nuclei and the Cosmic-Microwave-Background (CMB) photon during extragalactic propagation.}

Even though uncertainties are still persist, our attention focuses on scattering process of UHECR and UHE neutrino with {\it known} targets such as the nucleons in the atmosphere. We anticipate that such `fixed-target' experiment can provides a testing ground for particle physics beyond TeV scale~\cite{Anchordoqui:2018qom}. Indeed the collision energy can be as high as $\sqrt{m_N E_{\rm CR,\nu}}= \mathcal{O}(1-100)$ TeV given that the energy of the colliding cosmic ray  or neutrino  particle is around $\mathcal{O}(1)$ PeV$-\mathcal{O}(10)$ EeV with the nucleon mass $m_N\simeq 1$ GeV.

As a concrete example of relevant physics showing only above 1 TeV, we focus on the electroweak sphaleron process in this paper. The sphaleron is predicted to take place at around $10$ TeV in the standard model ~\cite{Manton:1983nd, Klinkhamer:1984di, Kuzmin:1985mm,Fukugita:1986hr, Shaposhnikov:1987tw, Ringwald:1988yt} but never has been experimentally tested and confirmed yet~\cite{Rubakov:1996vz,Ringwald:2002sw}. 
 We also study the microscopic blackholes predicted in low scale gravity scenarios for the hierarchy problem~\cite{Dimopoulos:2001hw,Giddings:2001bu}.\footnote{See \cite{Park:2012fe} for a recent review on microscopic blackhole and \cite{Arsene:2013ria,Arsene:2013nca,Arsene:2016kvf} for quantum blackholes.} The process of sphaleron and that of microscopic black hole share common properties such as large multiplicities and grown interaction cross sections at higher energies. We show that these common properties lead to observational consequences at the air-shower detector array experiments such as Telescope Array Experiment (TA) and Pierre-Auger Observatory (Auger)~\cite{Jezo:2014kla, Anchordoqui:2001ei, Anchordoqui:2001cg, Emparan:2001kf, Ringwald:2001vk, Feng:2001ib, Anchordoqui:2003jr, Ahn:2003qn, Anchordoqui:2005ey, Mack:2018fny}. { Also, black hole-induced UHE$\nu$-nucleon scattering in the IceCube detector has been studied \cite{Mack:2019bps}}.
 
 The paper is organized as follows: in Sec.~\ref{sec:cross sections}, we discuss the cross sections of electroweak sphaleron and microscopic black hole events.  In Sec.~\ref{sec:event rates}, we discuss the event rates of new physics processes taking the GZK neutrino flux. In Sec.~\ref{sec:features}, we show noticeable phenomenological features for each case and discuss potential detection of new physics effects. The recently reported `muon deficits' in hadronic interaction models~\cite{Aab:2014pza,Aab:2014dua,Abbasi:2018fkz} are also discussed. Finally, we conclude in Sec.~\ref{sec:conclusion}. The appendix includes all  details of the calculations used in the paper. 

%%%%%%%%%%%%%%%%%%%%%%%
\section{Production cross sections for sphaleron and black hole\label{sec:cross sections}}
%%%%%%%%%%%%%%%%%%%%%%%

In this section, we discuss phenomenological details of the air-showers induced by electroweak sphaleron and microscopic black hole. The air-showers typically accompany with large multiplicity in signals. The showers are boosted so that the final-state particles are highly collinear and confined within a small separation angle $\delta \theta \sim \mathcal{O}(1/\gamma_{\rm boost})$, where $\gamma_{\rm boost}=1/\sqrt{1-v^2}$ is the relativistic gamma factor of the produced particles with velocity $v$. 

%%%%%%%%%%%%%%%%%%%%%%%
\subsection{Parton level cross sections}
\label{subsec:sphaleron}
%%%%%%%%%%%%%%%%%%%%%%%

The electroweak sphaleron is predicted within the standard model as a  saddle point solution to the classical field equation of the electroweak gauge theory~\cite{Manton:1983nd,Klinkhamer:1984di}.   It is a highly unstable configuration thus is not directly observable. However its decay products, the Standard Model gauge bosons~\cite{Ringwald:1989ee} and fermions, leave observable effects \cite{Morris:1993wg} and reveal their presence in processes in the early universe. It also induces directly measurable signals in high energy collisions of { UHE neutrino} as we will closely study in this paper.

Sphaleron is involved in baryon number generation by inducing Chern-Simon (CS) number changing ($\Delta n_{\rm CS} = \pm n$) and baryon number ($B$) and lepton number ($L$) violating ($\Delta(B+L)\neq 0$, $\Delta(B-L)=0$) processes.  The generated baryon and lepton numbers are quantized as $\Delta B =\Delta L= \pm 3n$, where $n \in \mathbb{Z}$ is an integer number.   As the sphaleron process is effective in unbroken phase of electroweak symmetry, the baryon number, if generated before electroweak symmetry breaking, will be `wiped out' by sphaleron processes~\cite{Kuzmin:1985mm,Fukugita:1986hr}.  On the other hand, the generated lepton number is `converted' into baryon numbers via sphaleron. Therefore, it is important in baryogenesis from the lepton number generation~\cite{Fukugita:1986hr,Ringwald:1988yt}.  

Beside the role in baryogenesis, even though the sphaleron is robustly predicted {\it within} the Standard Model, it has never been experimentally tested {in the collider experiments,} due to the high sphaleron threshold at $E_{\rm sph}\approx 10$ TeV. The UHE neutrinos are particularly interesting since they reach this high threshold. According to recent works by Ellis and his collaborator~\cite{Ellis:2016ast,Ellis:2016dgb}, the production cross section of Sphaleron in scattering of $i$ and $j$ initial partons can be relatively large\footnote{See however,~\cite{Funakubo:2016xgd,Tye:2017hfv}. Conventionally it has been believed that the production cross section of a sphaleron in particle collision is exponentially suppressed even when the collision energy is beyond the threshold~\cite{Ringwald:1988yt}, $\hat{\sigma}_{ij \rightarrow {\rm Sphaleron}} \propto e^{-E/E_{\rm sph}}$.}:
\begin{eqnarray}
\hat{\sigma}_{ij \to {\rm EW Sph}}(E_{\rm CM}) & \simeq & \frac{p}{m_W^2} \theta(E_{\rm CM} - E_{\rm sph}), \nonumber \\
& \equiv & \frac{1}{\Lambda_{\rm Sph}^2} \theta(E_{\rm CM} - E_{\rm sph}) \label{eq:sph}
\end{eqnarray}
where $m_W =80.38$ GeV, $\theta(x)$ is the Heaviside step function and an unknown parameter $p \lsim 1$ encapsulates the unknown theoretical details. The cross section is not suppressed beyond the threshold because the sphaleron process is a collection of all possible processes over periodic vacua and the multiple contributions from all vacua overcome the exponential suppression factor~\cite{Tye:2015tva,Tye:2017hfv}. We take this cross section as a benchmark expression for the new physics `sphaleron'.

A microscopic black hole forms through collision of particles with a relatively low energy, $\sqrt{s}\gsim 1$ TeV, in low-energy gravity scenarios~\cite{Dimopoulos:2001hw,Giddings:2001bu}. Once produced, a microscopic black hole may explosively decay into multiple number of photons and also other Standard Model particles through Hawking radiation with high Hawking temperature, $T_{\rm Hawking} > m_t$~\cite{Hawking:1974rv,Hawking:1974sw}. Thus the collision of {UHE neutrino} with a nucleon in the atmosphere is a potential source of a high-multiplicity event.

The parton level production cross section of ($i+j \rightarrow {\rm black \ hole} +X$) is approximately given as 
\begin{eqnarray}
\hat{\sigma}_{ij \to \rm BH} (E_{\rm CM}) & \approx & \pi \left(G_D E_{\rm CM}\right)^{\frac{2}{D-3}}, 
\label{eq:bh}
\end{eqnarray}
where $G_D = 1/M_D^{D-2}$ is the gravitational constant in $D(=4+n)$-dimensions with $n$-extra compact dimensions~\cite{Dimopoulos:2001hw,Giddings:2001bu,Park:2001xc,Yoshino:2002tx}. The Schwarzschild radius is rapidly growing as $r_{\rm Sch} \approx (G_D E)^{1/D-3}$ so that the resultant cross section grows too. Assuming $M_D\sim {\rm TeV}$ in low energy gravity scenarios, the cross section can be sizable as $\hat{\sigma} \gsim 1/{\rm TeV}^2$ when $E\gsim {\rm TeV}$.

%%%%%%%%%%%%%%%%%%%%%%%%%%%%%%%%%
\begin{figure}[t]
{\includegraphics[width=.45\textwidth]{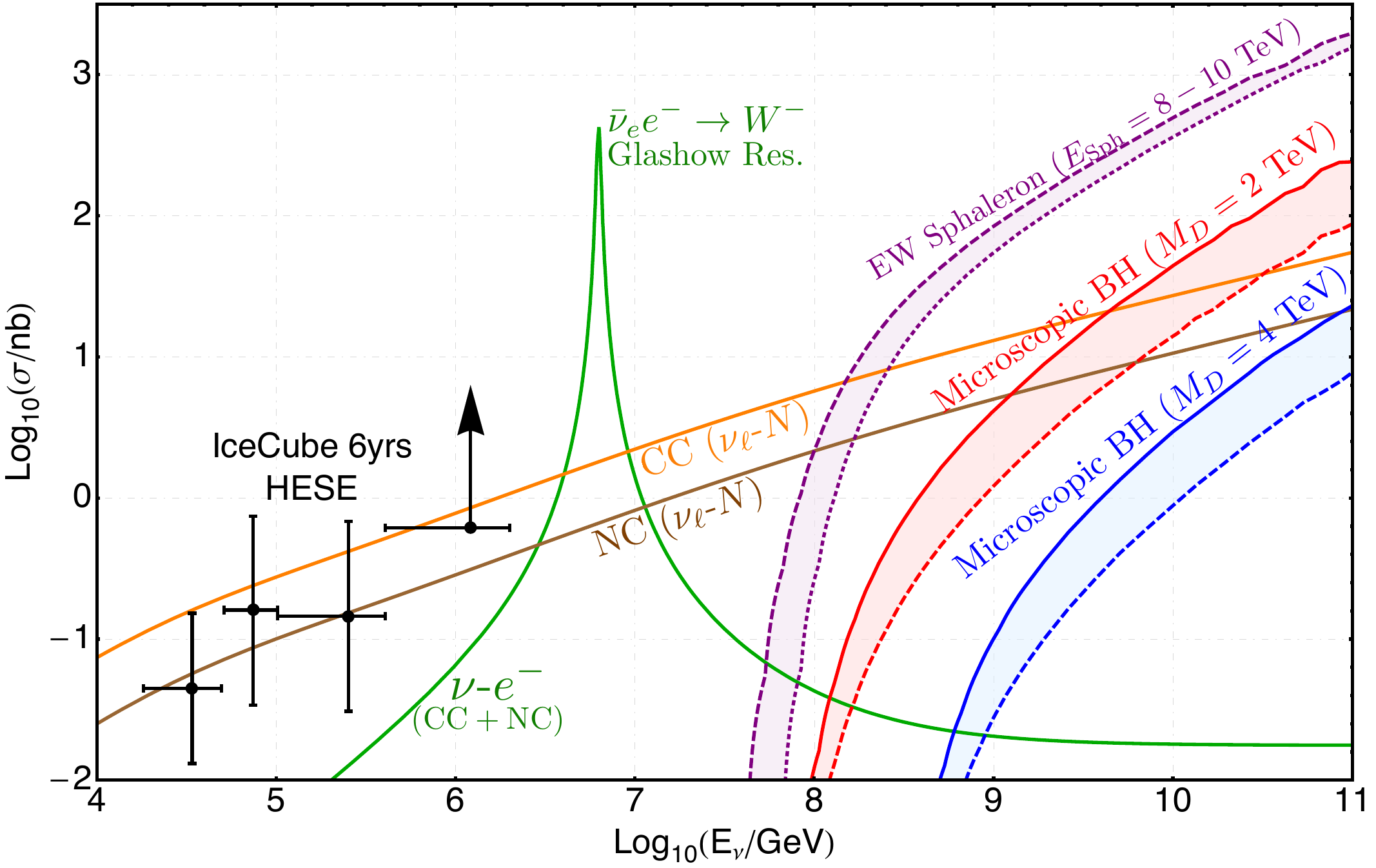}} \,
\caption{{\bf [Neutrino-induced cross sections with nucleon or electron.]}
The CC/NC neutrino-nucleon deep-inelastic scattering (DIS) (orange and brown), Glashow resonance (green), electroweak sphaleron with $E_{\rm Sph} = 8-10$ TeV (purple band), microscopic blackholes with $M_D = 1$ TeV (red band) and $M_D = 2$ TeV (blue band) for the number of compact dimensions, from $n = 2$ (bottom) to $n = 6$ (top). We fixed the ratio between the minimum energy for black hole production and the gravity scale as $x_{\min} \equiv M_{\min}/M_D = 5$ for all cases. The black dots with error bars are $\nu-N$ cross section obtained from the 6-years IceCube high-energy starting events (HESE) shower data set.}
\label{UHECrossSection}
\end{figure}
%%%%%%%%%%%%%%%%%%%%%%%%%%%%%%%%%

%%%%%%%%%%%%%%%%%%%%%%%
\subsection{$\nu-N$ cross section}
%%%%%%%%%%%%%%%%%%%%%%%

The $\nu-N$ cross section can be obtained by taking the parton distribution functions (PDF), $f_q(x, \hat{Q}^2)$,  for a quark, $q$,  in nucleon, $N$. The total cross sections for electroweak sphaleron and microscopic black hole are respectively given after the PDF convolution for nucleon:
\begin{eqnarray}
& & \sigma_{\rm EW Sph, BH}^{\nu N} (E_{\rm lab}) \nonumber \\
& & = \int_{x_{\min}}^1 dx \ f_q (x, 2x m_N E_{\rm lab}) \ \hat{\sigma}_{\rm EW Sph, BH}^{\nu q} (\hat{s}),
\end{eqnarray}
where $\hat{s}=2x m_N E_{\rm lab}$.  The parton level cross sections are given in Eq.~\ref{eq:sph} and Eq.~\ref{eq:bh}, respectively. The input parameters are $m_N$  (the mass of Nucleon), $E_{\rm lab}$ (the collision energy in lab frame) and $\hat{s}$ (the collision energy at the parton level). The minimum energy for making a black hole (sphaleron) is controlled by $x_{\rm min}={\rm Min}[\hat{s}/(2m_N E_{\rm lab})]$. It is noted that  if we set the unknown parameter of the sphaleron cross section  $p \sim \mathcal{O}(1)$, the  values of cross sections for sphaleron and black hole  are numerically close to each other  with $M_{{\rm D}}\sim \mathcal{O}(1)$ TeV.  Even though the production cross section for black holes eventually overtakes the sphaleron production, the luminosity of {UHE neutrino} becomes smaller at higher energies therefore the total event rates stay similar for the two different cases.

In Fig.~\ref{UHECrossSection}, we  show the relative sizes of the standard model NC/CC cross sections \cite{Gandhi:1998ri} and the expected cross sections from sphaleron and black hole with various parameter choices. We use {\tt NNPDF 3.1} \cite{Ball:2017nwa} as the parton distribution function in the evaluation for CC/NC, sphaleron \cite{Ellis:2016dgb}, and black hole production processes. We find that the sphaleron and black hole interactions will become increasingly important at higher energies as we expected: at low energies below $10^{6-7}$ GeV the standard model CC/NC neutrino-nucleon deep-inelastic scattering (DIS) dominates over other interactions as depicted by orange and brown lines. In a narrow resonance region at the $W$-boson threshold  at $E_\nu = m_W^2 / 2 m_e \simeq 6.3$ PeV, the $W$ production process,  $\bar{\nu}_e e^- \rightarrow W^-$,  dominates as depicted by the green, mountain shape line (Glashow resonance). Above $\mathcal{O}(100)$ PeV,  the new high multiplicity events, induced  by electroweak sphaleron and black hole, become important and eventually dominate over the standard model CC/NC interactions.\footnote{CC/NC interaction cross section has dominant contributions at small parton fraction as $x \lsim 10^{-4}$ and cross section increase as $\sigma \sim E^{\delta}$ where the parton distribution fitted $x f_q (x) \sim x^{-\delta}$. In contrast, new physics (black hole or sphaleron) cross sections come from only large $x$ region as $x \gsim x_{\min} = E_{\min}^2/2m_N E_\nu \gsim 10^{-4}$.} We depict the sphaleron events and black events with the parameters: $E_{\rm Sph} = 8-10$ TeV for sphaleron (purple band) and $M_D = 1-2$ TeV with fixed $x_{\min} \equiv  M_{\min}/M_D = 5$ for black hole. We take $M_D =1$ TeV, $M_{\min} = 5$ TeV (red band) and $M_D = 2$ TeV, $M_{\min} = 10$ TeV (blue band) as our benchmark parameter choices. The bands for black hole are for various numbers of extra dimensions $2 \leq n \leq 6$ from bottom ($n = 2$) to top ($n = 6$). Finally  the observational results for neutrino-nucleon cross section are depicted by black dots with error bars, which are from the 6-years IceCube data. In particular, we take the High Energy Starting Events (HESE) \cite{Aartsen:2017kpd,Bustamante:2017xuy}, whose starting points of cascade or track are located inside the IceCube detector.

The proton-proton and proton-air cross sections are large and the QCD showers are dominant in low $X_0$ region. However, we show that  the QCD background can be greatly reduced even below the new physics level by imposing a cut in large \sout{$X_0^{\rm cut}$} atmospheric depth of injection position. The details are discussed later in Section \ref{subsec:X0cut}. After the cut, we take the  sphaleron-induced and black hole air-showers with neutrino-nucleon collision as our signals, and CC/NC with neutrino-nucleon collision as our backgrounds.

%%%%%%%%%%%%%%%%%%%%%%%
\section{Event rates and limits on high-multiplicity processes \label{sec:event rates}}
%%%%%%%%%%%%%%%%%%%%%%%

\begin{figure}[t]
\centering
\includegraphics[width=.45\textwidth]{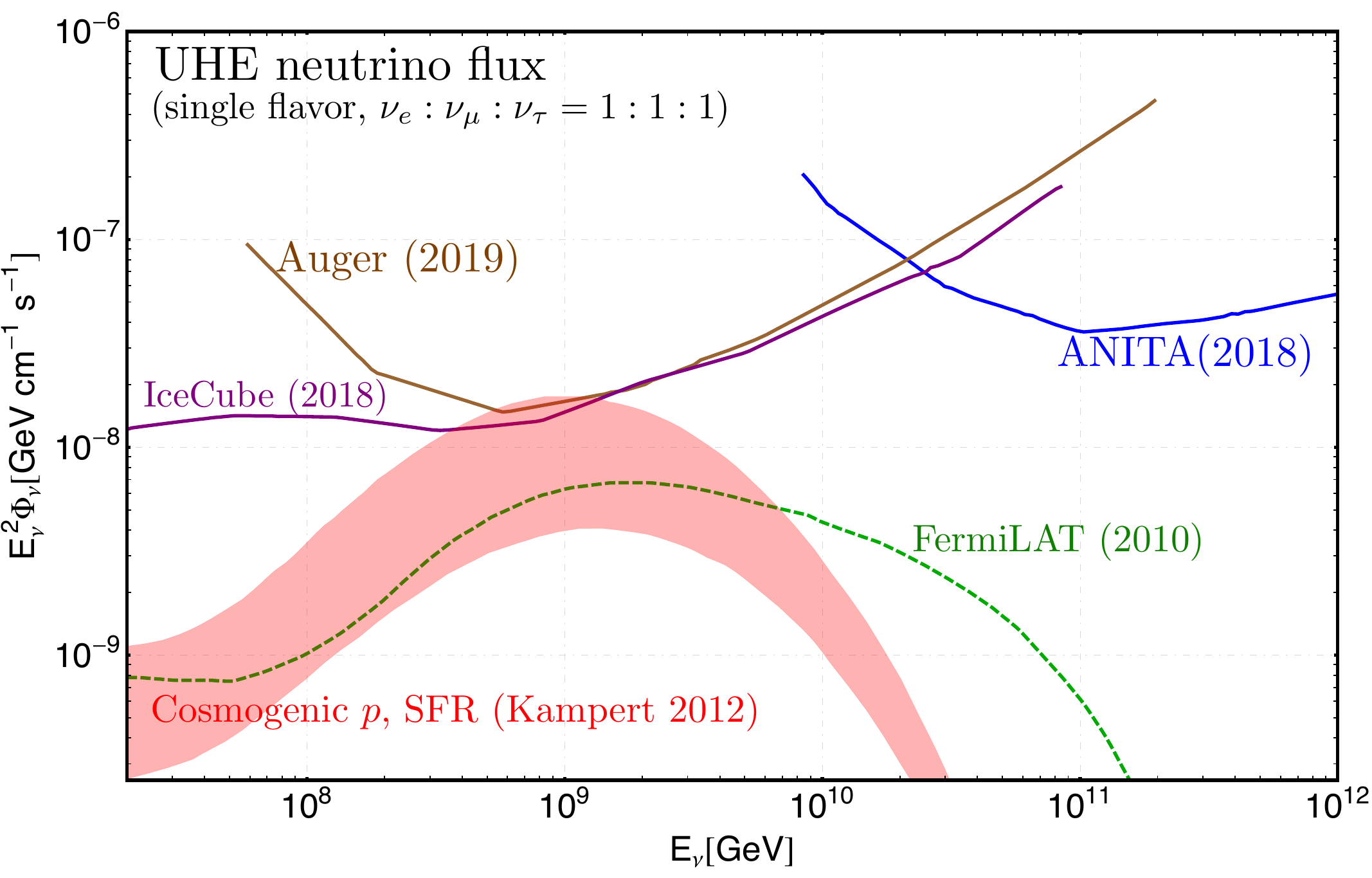}
\caption{{\bf [$E_\nu^2 \Phi_\nu$, GZK neutrino fluxes] Expected GZK neutrino fluxes and current experimental bounds.]} Expected GZK neutrino fluxes from the diffuse gamma-ray observation of Fermi-LAT (green, dashed) \cite{Ahlers:2010fw} and SFR source evolution (red band) \cite{Kampert:2012mx}. The current experimental bounds from IceCube (purple) \cite{Aartsen:2018vtx}, ANITA (blue) \cite{Allison:2018cxu} and Auger (brown) \cite{Aab:2019auo}.} \label{GZKneutrinoFluxBounds}
\end{figure}

As the fluxes of diffuse gamma rays and the GZK neutrino are correlated, the flux of GZK neutrinos can be determined from the diffuse gamma ray data taken e.g. by Fermi-LAT~\cite{Ahlers:2010fw}.
In  Fig. \ref{GZKneutrinoFluxBounds}, we collect the UHE neutrino fluxes obtained from various observational sources in the range of the energy, $E_{\rm sh} \subset (10^8, \ 10^{10})$ GeV~\cite{Aartsen:2015zva,Ishihara:2016pty}. In particular, we show the upper bounds on the fluxes obtained from `direct measurements' from IceCube (2008-2014) \cite{Ishihara:2016pty} and Pierre Auger (2004-2013) \cite{Aab:2015kma}. We also show the flux obtained from `indirect measurement' of Fermi-LAT gamma ray data \cite{Ahlers:2010fw} with different minimum energies of extragalactic cosmic rays involved in the photopion production. 

\begin{figure*}[t]
\centering
{\includegraphics[width=0.335\textwidth]{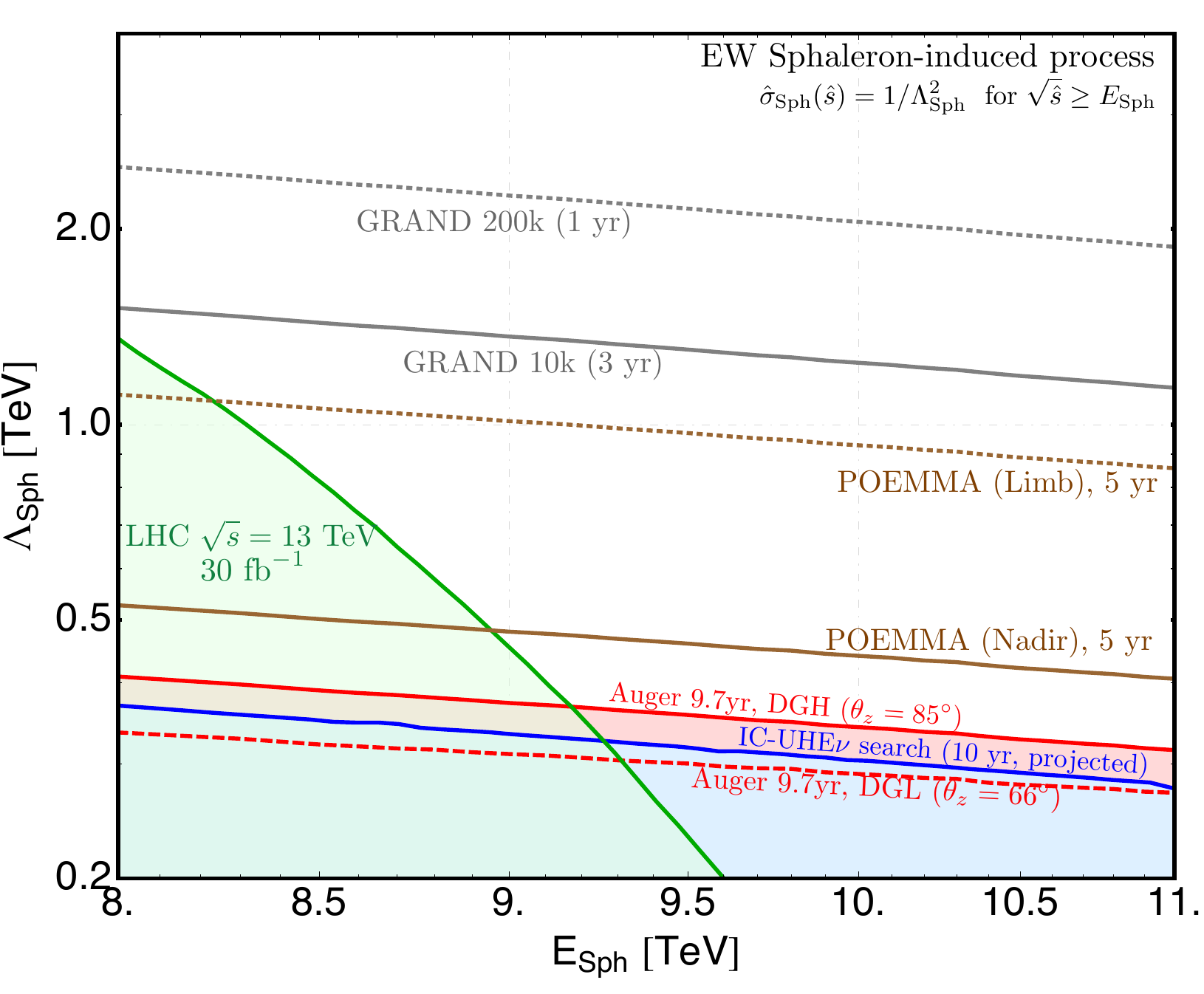}} 
{\includegraphics[width=0.325\textwidth]{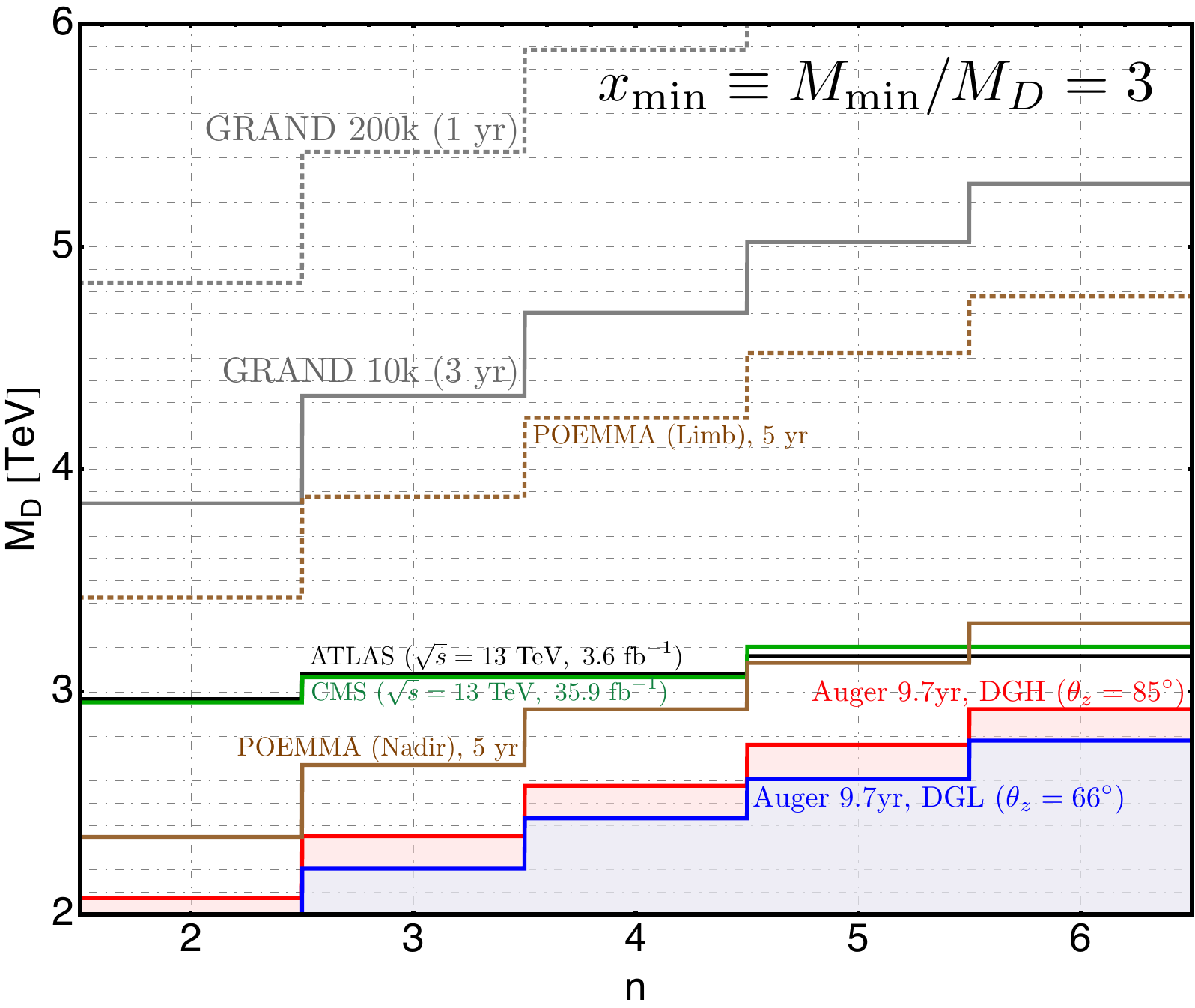}} 
{\includegraphics[width=0.325\textwidth]{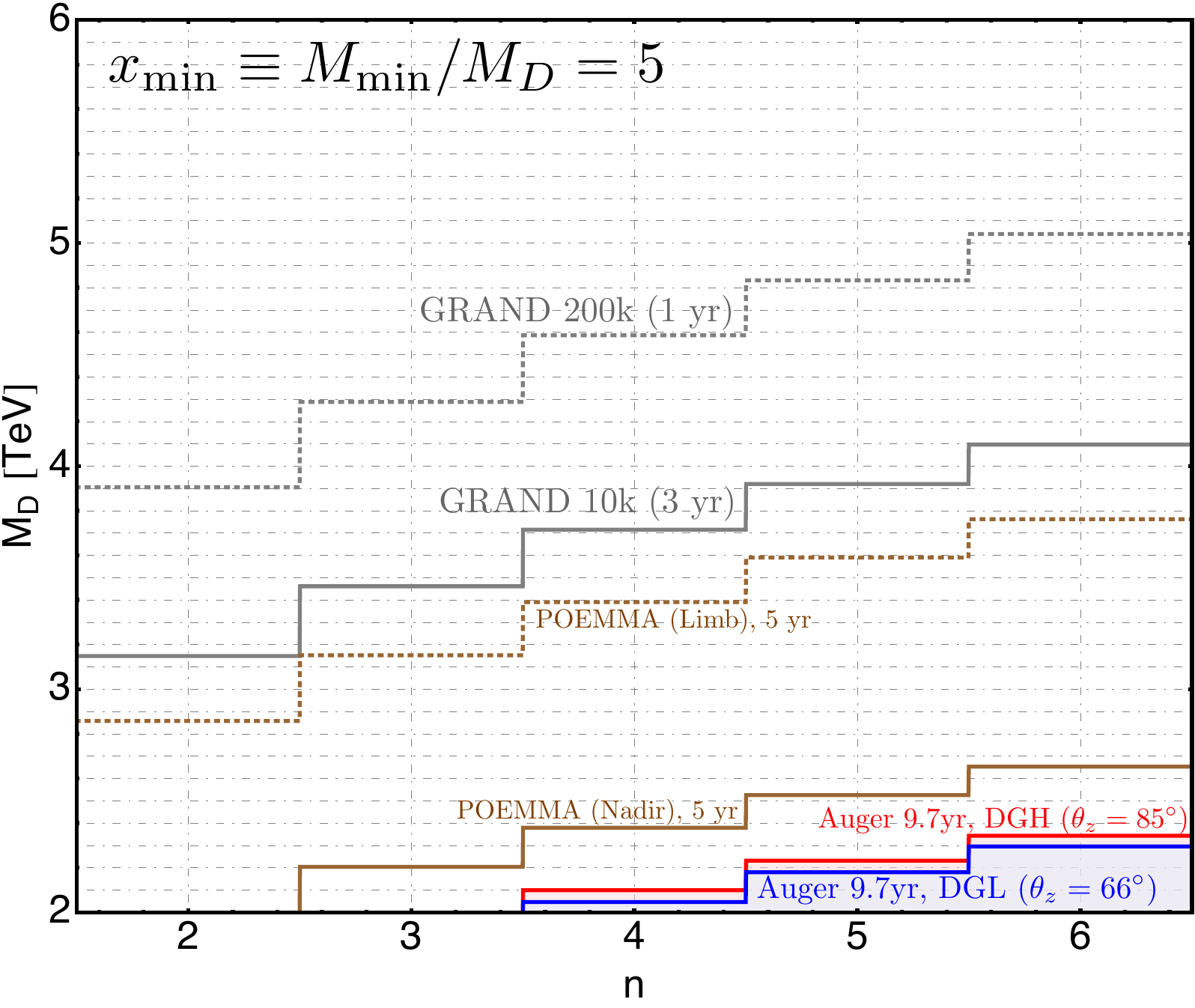}} 
 \caption{90$\%$ confidence level lower limits of cross section scale for electroweak sphaleron $\Lambda_{\rm Sph}$ (Left Panel) and the extra-dimensional model parameters ($n,M_D$) for $x_{\min} \equiv M_{\min}/M_D = 3$ (Middle Panel) and $x_{\min} = 5$ (Right Panel). For Auger limits, we use effective area for Down-going high zenith ($\theta_z = 85^\circ$, `DGH') and Down-going low zenith ($\theta_z = 66^\circ$, `DGL') UHE neutrino searches \cite{Aab:2019dav, Aab:2019ogu}. We show the current excluded regions by CMS \cite{Sirunyan:2018xwt}, ATLAS \cite{Aad:2015mzg, Aaboud:2016ewt} and IceCube \cite{Ishihara:2016pty}. The expected lower limits in future experiments (GRAND \cite{Alvarez-Muniz:2018bhp} and POEMMA \cite{Olinto:2020oky}) with several choices of total exposures for UHE$\nu$ search, also are shown.}
 \label{Limits}
\end{figure*}

{
With given expectations on UHE neutrino fluxes in Fig.~\ref{GZKneutrinoFluxBounds}, we obtain lower limits of parameters for electroweak sphaleron ($\Lambda_{\rm Sph}$) and extra-dimensions ($n$, $M_D$ and $M_{\min}$) in Fig.~\ref{Limits}. Current observatories such as Auger \cite{Aab:2019dav, Aab:2019ogu} and IceCube \cite{Ishihara:2016pty}, and also future proposed facilities (e.g., GRAND \cite{Alvarez-Muniz:2018bhp} and POEMMA \cite{Olinto:2020oky}) are considered in order to provide limits. For the evaluation of both signal (sphaleron, black hole) and background (electroweak CC/NC) event rates, we follows the expressions in Appendix.~\ref{EvRateAirshowerCCNCNP}. We use the UHE neutrino flux estimated from Ref.~\cite{Ahlers:2010fw} (green dashed line in Fig.~\ref{GZKneutrinoFluxBounds}) as our benchmark choice. For the discrimination between signal and background air-showers, we explain the details in Section~\ref{airshower}.

On the other hand, CMS \cite{Sirunyan:2018xwt} and ATLAS \cite{Aad:2015mzg, Aaboud:2016ewt} experiments provide lower limits on the same parameters ($\Lambda_{\rm Sph}$, $n$, $M_D$ and $M_{\min}$) from the analysis of LHC Run 2 data. However, as shown in Fig.~\ref{Limits}, current constraints from LHC are only sensitive for $\sqrt{\hat{s}} \lsim 10$ TeV. Typical searches of black hole production only can give exclusion limits for $1 < x_{\min} \lsim 3$, where the semi-classical description of evaporation is not fully guaranteed \cite{Park:2011je}. In contrast, due to the large (as $\sqrt{s} \simeq 130$ TeV for $E_\nu \simeq 10^{10}$ eV) center-of-mass of expected UHE$\nu$-nucleon collision in ground array observatories, astrophysical searches can be more relevant for larger $x_{\min}$, up to $x_{\min} \lsim 5$, as shown in the right panel of Fig.~\ref{Limits}.
}

 %%%%%%%%%%%%%%%%%%%%%%%%%%%%%%%%%%%%%%%%%%%
%%%%%%%%%%%%%%%%%%%%%%%%%%%%%%%%%%%%%%%%%%%

%%%%%%%%%%%%%%%%%%%%%%%%%%%%%%%%%%%%%%%%%%%
%%%%%%%%%%%%%%%%%%%%%%%%%%%%%%%%%%%%%%%%%%%

%%%%%%%%%%%%%%%%%%%%%%%%%%%%%%%%
\section{Features of New Physics events \label{sec:features}}
%%%%%%%%%%%%%%%%%%%%%%%%%%%%%%%%

 In this section, we describe some evident features of new physics events based on the symmetry principles and the quantum nature of Hawking radiation, which eventually provide useful guidelines to single out the new physics events from the background events.

%%%%%%%%%%%%%%%%%%%%%%%%%%%%% 
\subsection{Signals from Sphaleron: $(B-L)$ symmetry}
%%%%%%%%%%%%%%%%%%%%%%%%%%%%%

For sphaleron, due to the symmetries of baryon number and lepton number, $\Delta (B-L)=0$  and $\Delta (B+L) =3n$ with an integer number $n$, the particle contents of the sphaleron process in the final state are almost uniquely determined. 
For instance, $\nu_e-N$ collision generates $(10~{\rm fermions})+ (n_B~{\rm bosons})$ particles satisfying  $\Delta B =\Delta L = -3$. 
The number of bosons $(n_B~{\rm bosons})$ is the sum of the number of gauge bosons ($n_W+n_Z$) and the number of Higgs bosons ($n_H$). 

More precisely, if up-type quark is in the initial nucleon $N$, the final state is
\begin{eqnarray}
N(u) + \nu_e \rightarrow L + Q + n_W W + n_Z Z + n_H H
\end{eqnarray}
where $L$ ($Q) $ stands for the primary leptons (quarks), respectively. For example, $L = \mu^+ + \bar{\nu}_\tau$ and $Q = \bar{t} + 2 \bar{b} + 2 \bar{c} + \bar{s} + \bar{u} + \bar{d}$ is one of the possible minimal choices satisfying $L=1$ to $L=-2$ and $B=1/3$ to $B=-8/3$, respectively. In general, there are 2 antileptons ($L=-2$) and 8 antiquarks ($B=-8/3$) in a final state with possible addition of electroweak bosons. The secondary leptons from the decay of primary (heavy) quarks and gauge bosons are distinguishable as they are less energetic compared to the primary leptons.  

Taking all the properties discussed above, we summarize the rules for the signal configuration:
\begin{itemize}
\item total ${\rm SU(2)}_L$ isospin is conserved, 
\item ${\rm SU(3)}$ color is conserved and the total color is singlet if initial and final states are considered all together, 
\item $\Delta B_i = \Delta L_j = \Delta N_{\rm CS}$ for each families. \\ ($i,j=1,2,3$)
\item total $U(1)_{\rm EM}$ and $B-L$ charge are conserved.
\end{itemize}

%%%%%%%%%%%%%%%%%%%%%%%%%%%%%%%%%%%%%%%%%%%
\begin{figure}[h]
\centering
\includegraphics[width=.48\textwidth]{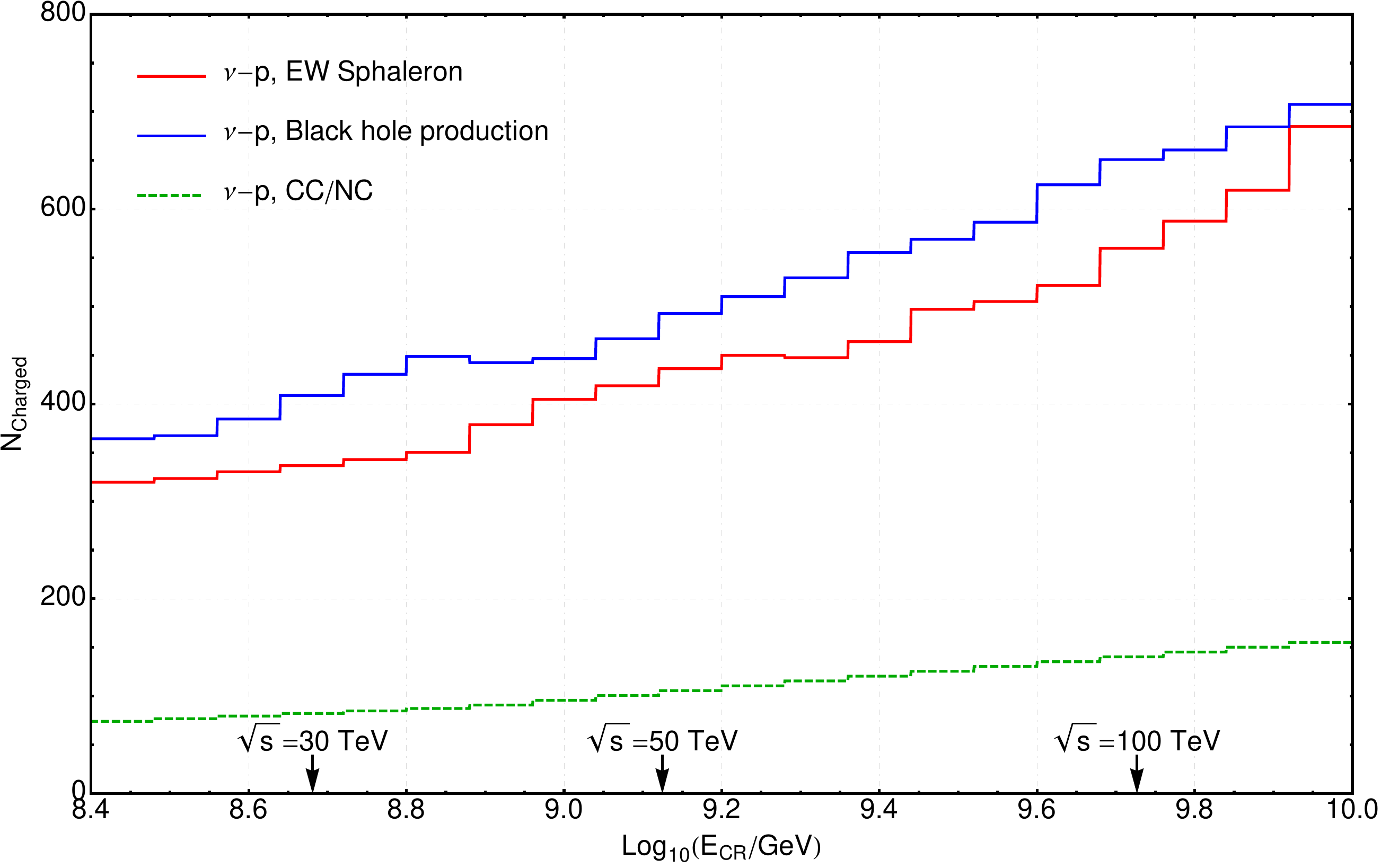}
\caption{{\bf [Charged particle multiplicity $N_{\rm Charged}$ ($\simeq N_{\pi^\pm}^{(0)}$)].} The average number of charged particles after the primary hard interactions of neutrino-proton collision with CC/NC (green dashed line), EW sphaleron (red solid line) and microscopic black holes productions (blue solid line).
\label{Charged_Particle_multiplicity_primary_int}}
\end{figure}
%%%%%%%%%%%%%%%%%%%%%%%%%%%%%%%%%%%%%%%%%%%

Typically, the final state of the sphaleron-induced process consists of $\mathcal{O}(10)$ hadronic jets and a few additional leptons in high energy domain and each particle carries an energy of about $E \approx  M/n_{\rm primary}$, where $M$ is the new physics scale and $n_{\rm primary}$ is the number of primary decay products. The high multiplicity of hadronic components  leads to  lower individual pion's energy $E_\pi\approx E_{\rm CR}/(N_{\pi^\pm} + N_{\pi^0})$, thus the amount of energy loss is relatively smaller before reaching the critical energy, $E_{\rm crit} =(1-10)$ GeV  \cite{Matthews:2005sd}.
In Fig.~\ref{Charged_Particle_multiplicity_primary_int} we show the number of charged particles (charged particle multiplicity, $N_{\rm Charged}\simeq N_{\pi^\pm}^{(0)}$) from the different origins. The new physics induced air-showers have  larger $N_{\rm Charged}$. We used hadronic interaction models, {\tt QGSJET II-04} \cite{Ostapchenko:2010vb} and {\tt EPOS LHC} \cite{Pierog:2013ria} for calculation.

Below the critical energy, the charged pion's decay length, $\gamma c \tau_{\pi^\pm}$, becomes short in comparison to the interaction depth $\lambda_{\pi} \simeq 120$ g cm${}^{-2}$ of the atmosphere. In this stage, the charged pions decay mainly via $\pi^+ \rightarrow \mu^+ \nu_\mu$ before they interact with nuclei in air molecules.
 To some extent, the high multiplicity processes of new physics are similar to the processes due to heavy nuclei in the sense of the superposition model \cite{Engel:1992vf}. The extensive air shower from a heavy nuclei of the atomic mass $A_{\rm CR}$ with primary energy $E_{\rm CR}$ can be considered as a parallel copy of $A_{\rm CR}$ proton air-showers, and each of the proton carries the primary energy of about $E_{\rm CR}/A_{\rm CR}$. It is important to note that the number of muons included in the air shower at the observation level $h_{\rm obs}$ scales as $N_\mu \propto (A_{\rm CR})^{1-\beta}$ \cite{Matthews:2005sd} where $\beta = \ln \frac{N_{\pi^\pm}}{\frac{3}{2} N_{\pi^\pm}} = 0.85$ and $(A_{\rm CR})^{1-\beta} \simeq 1.8$ for the iron nuclei, i.e. $A_{\rm CR} = 56$. 

\begin{figure*}[ht]
\centering
\includegraphics[width=.8\textwidth]{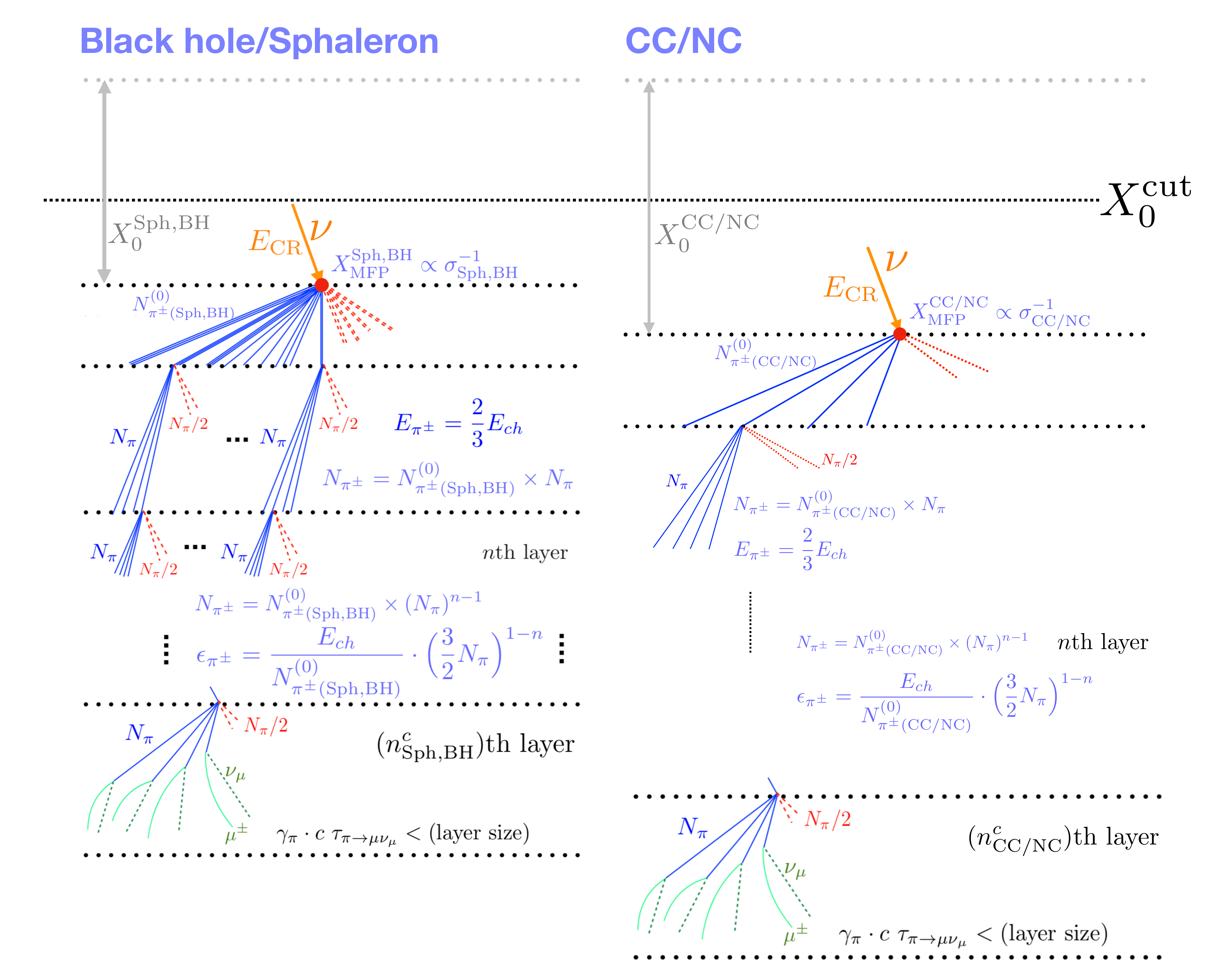}
\caption{{\bf [Schematic view of typical extensive air showers for two cases of primary hard interactions between neutrino and nucleon.]} Left: EW sphaleron or black hole productions with a primary neutrino and a target nucleon. Right: CC/NC interactions with a primary neutrino and a target nucleon. Note that QCD backgrounds are removed by $X_0$ cut. Initial charged pion multiplicity $N_{\pi^\pm({\rm Sph,BH})}^{(0)}$, and $N_{\pi^\pm({\rm CC/NC})}^{(0)}$ are shown in Fig. \ref{Charged_Particle_multiplicity_primary_int}.\label{EAS_view_schematic}}
\end{figure*}

As a  summary, the main features of air-showers from the different origins are collectively shown in Fig. \ref{EAS_view_schematic}: the schematic picture of each air-shower event from different primary interactions { (sphaleron, black hole and CC/NC)}.

%%%%%%%%%%%%%%%%%%%%%%%%%%%%%%%%%%%%%%%%%%%
%%%%%%%%%%%%%%%%%%%%%%%%%%%%%%%%%%%%%%%%%%%

%%%%%%%%%%%%%%%%%%%%%%%
\subsection{New Physics air-shower features at ground arrays}\label{airshower}
%%%%%%%%%%%%%%%%%%%%%%%

In addition to the high multiplicity of energetic hadronic components ($n_j\gsim 10$) with associated leptons, we also notice distinguishable features of the new physics events by performing realistic simulation of the air-shower events.  
 The simulation is carried out aiming to see the new physics effects in the energy range $10^{15}$ eV $-$ $10^{20}$ eV or collision energy of $\sqrt{s} = 10 - 300$ TeV).  As a benchmark, we set $E_{\rm Sph} = 9.0$ TeV for electroweak sphaleron-induced events. In particular, we consider minimal signals without additional $W$-boson attached. From the events from the microscopic blackholes, we set $M_D = 1-2$ TeV and $M_{\rm min} = 5 M_D$, which is complimentary to the LHC searches~\cite{Aad:2015mzg,Aaboud:2016ewt}.
Several MC tools are used for generating the extensive air-shower events: {\tt BlackMax} \cite{Dai:2007ki} for the parton level black hole production and {\tt PYTHIA8} \cite{Sjostrand:2007gs} for primary parton shower and hadronization. Finally {\tt CORSIKA} \cite{CORSIKA} is used for extensive air-shower cascade where {\tt GHEISHA} at low energies and {\tt QGSJET II-04} \cite{Ostapchenko:2010vb} (or {\tt EPOS LHC} \cite{Pierog:2013ria}) at high energies are attached for hadronic interactions in air-shower cascade simulations. A particularly useful quantity is  the atmospheric slant interaction depth, $X(h)$, which is defined as the integrated density of atmosphere along the path of air-shower,
\begin{eqnarray}
X (h) & \equiv & \int_h^\infty dh' \ \rho_{\rm atm} (h')
\end{eqnarray}
where $\rho_{\rm atm}(h)$ is the density of the atmosphere at height $h$.

%%%%%%%%%%%%%%%%%%%%%%%%%%%%%%%%%%%%%%%%%%%
\begin{figure*}[t]
\centering
{\includegraphics[width=0.48\textwidth]{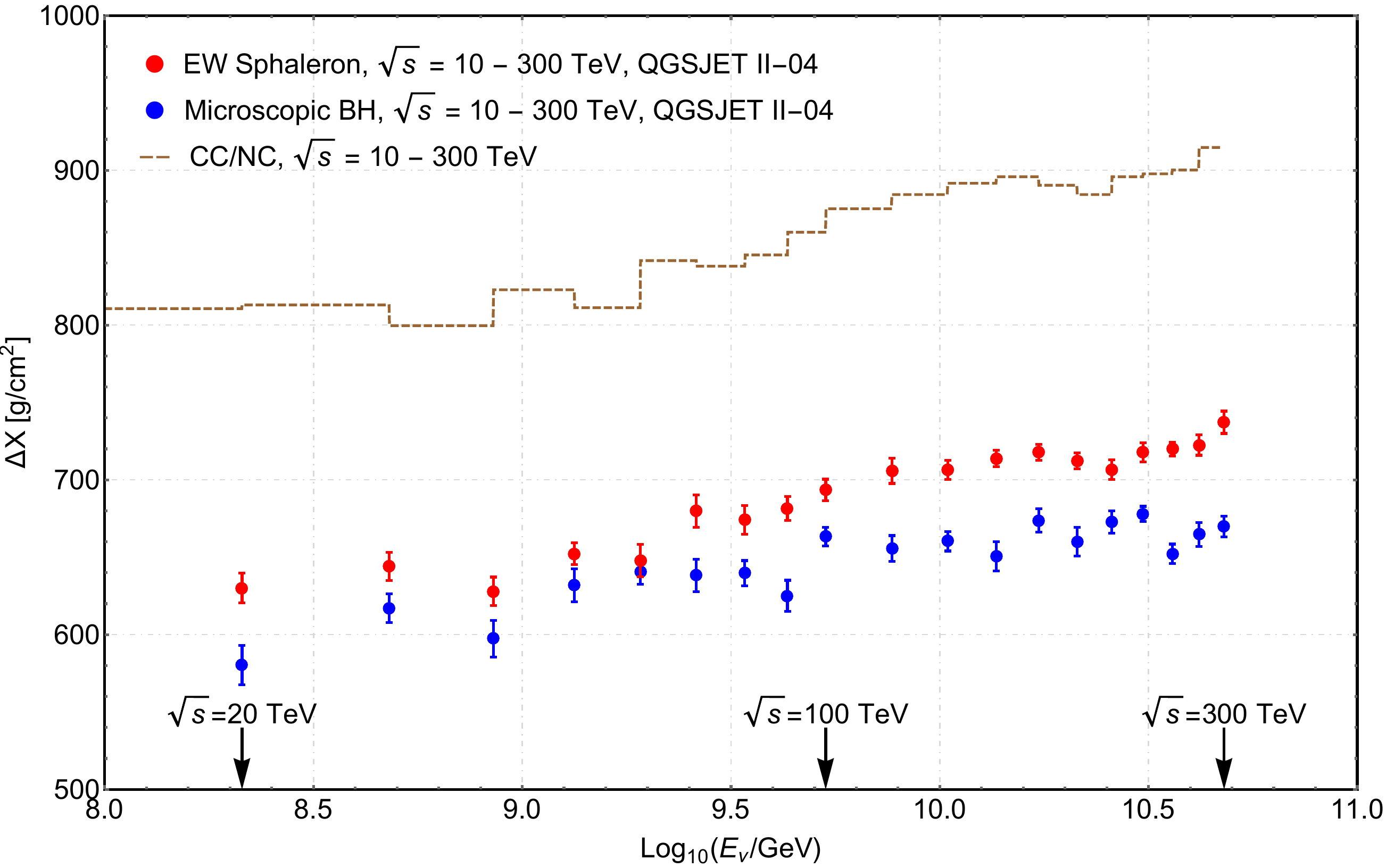}} \, 
{\includegraphics[width=0.48\textwidth]{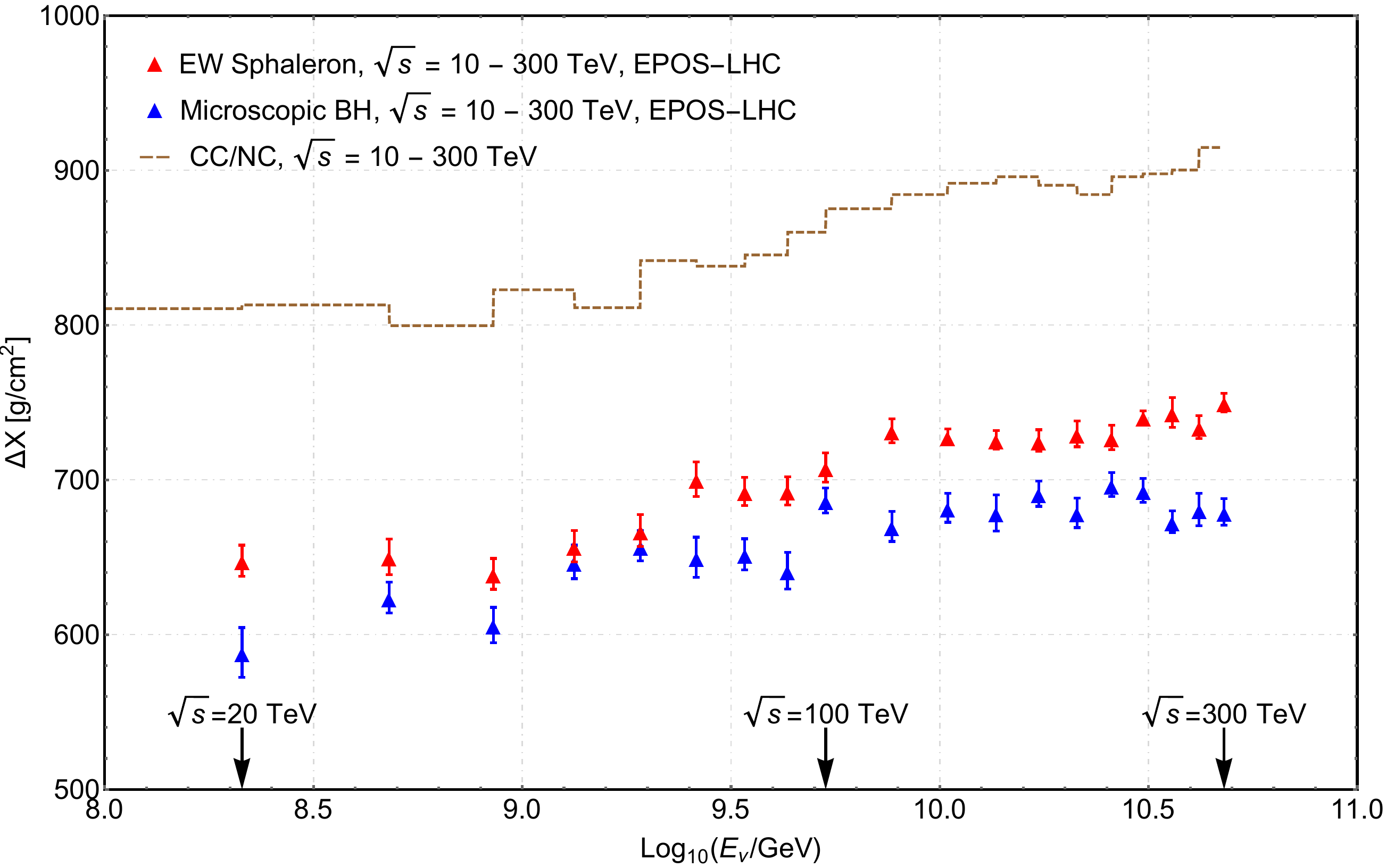}} 
 \caption{{\bf [The Monte Carlo simulation of expectations of $\Delta X$ for EW sphaleron (red dot), black hole (blue), and CC/NC (brown dashed line) in $E_{\rm CR} = 10^{8} - 10^{11}$ GeV.]} Note that circles (in the left panel) are the result with {\tt QGSJET II-04} \cite{Ostapchenko:2010vb} and triangles (in the right panel) are with {\tt EPOS-LHC} \cite{Pierog:2013ria} in its extensive shower event generation. $\Delta X$ values for CC/NC showers are also shown for comparison.}
 \label{LongiDistDeltaX}
\end{figure*}
%%%%%%%%%%%%%%%%%%%%%%%%%%%%%%%%%%%%%%%%%%%

We mainly focus on the observables listed below.
\begin{itemize}
\item $X_0$ (Fig. \ref{NevAfterX0cut}): The first interaction point of air-showers, defined as
\begin{eqnarray}
X_0 & \equiv & \int_{h_{\rm inj}}^\infty dh' \ \rho_{\rm atm} (h')
\end{eqnarray} 
where $h_{\rm inj}$ is the height of the starting point (or injection point) of the air-shower induced by the collision of the { UHE neutrino} and the nucleon in the atmosphere. The typical air-shower events induced by the standard QCD interactions have a relatively steep distribution but the new physics interactions with a smaller cross section, $\sigma_{\rm NP} \ll \sigma_{\rm QCD}$, induce much broader distribution in $X_0$ (See Fig. \ref{NevAfterX0cut}) as the probability distribution of the interaction point is given as
\begin{eqnarray}
P(X_0) \propto \exp \left( - \sigma_{\rm int} N_A A_{\rm atm}^{-1} X_0\right)
\end{eqnarray}
where $\sigma_{\rm int}$ is the cross section of primary interaction and $A_{\rm atm} = 14$ is the atomic mass of the atmosphere. 

\item $\Delta X (\equiv X_{\max} - X_0$) (Fig.~\ref{LongiDistDeltaX}): $\Delta X$ is the difference between the interaction depth position of the maximum charged particle multiplicity, $X_{\max}$, and the first interaction depth. In this work, we focus on $\Delta X$ because it is highly sensitive to the types of the relevant interactions. $\Delta X$ is observable at 24 fluorescence detector (FD) telescopes in the Auger observatories{, for $\theta_{\rm zenith} \lsim 60^\circ$, by measuring fluorescence light emitted from excited atmospheric (nitrogen) molecules in the range of $300-430$ nm \cite{Abraham:2009pm}. In addition, Auger can be sensitive to $X_{\rm max}$ measurement for highly-inclined air-showers ($\theta_{\rm zenith} = 60^\circ - 84^\circ$) by utilizing radio array facility \cite{Aab:2018ytv}. Measuring $X_{\max}$ and $\Delta X$ for inclined showers ($\theta_{\rm zenith} \gsim 50^\circ$) will be crucial to distinguish sphaleron/black hole signal showers from CC/NC background showers. Nevertheless, detecting the precise value of $\Delta X$ is also important in future searches. GRAND and POEMMA experiments are expected to be sensitive for $X_{\max}$ and $\Delta X$ with the resolution at the level of $\sim 20-40$ g/cm$^2$ \cite{Alvarez-Muniz:2018bhp, Olinto:2020oky}, which is beneficial for giving future limits on sphaleron/black hole showers.}

\end{itemize}

In Fig. \ref{LongiDistDeltaX}, we depict the Monte Carlo simulation result of $\Delta X$ for EW sphaleron (left, red dots with error bars) and black hole (right, blue dots with error bars) in $E_{\rm CR} = 10^{8} - 10^{11}$ GeV. All error bars indicate the statistical error of the mean value, which is root-mean-square divided by square-root of the number of entries in each energy bin. {Fig.~\ref{LongiDistDeltaX} indicates $\Delta X$ values for new physics showers and $\Delta X$ for ordinary CC/NC showers are shown in comparison.\footnote{$X_{\max}( = X_0 + \Delta X)$ distribution for new physics shower will be much broader since $X_0$ are almost uniformly distributed.} We can easily notice that the sphaleron and black hole induced air-showers have smaller $\Delta X (= X_{\rm max} - X_0)$ values compared to CC/NC air-showers. Our results are consistent with earlier simulation for sphaleron-induced \cite{Brooijmans:2016lfv} and black hole-induced \cite{Ahn:2005bi} air-showers. The new physics induced air-showers by sphaleron and black hole have broader distributions in their injection positions $h_{\rm inj}$.  (See Appendix. \ref{MFPavgMuonNumber} for more details).

Imposing $X_0^{\rm cut}$ (Fig. \ref{NevAfterX0cut}) and $\Delta X^{\rm cut}$ (Fig. \ref{LongiDistDeltaX}), our selection criterion for signal is given by
\begin{eqnarray}
i) \ X_0 & > & X_0^{\rm cut} = 1200{\rm g/cm}^2, \\
ii) \ \Delta X & < & \Delta X^{\rm cut} \simeq 50 \ {\rm Log}_{10} (E_{\rm CR}/{\rm GeV}) + 260{\rm g/cm}^2. \nonumber \\
\end{eqnarray}
One can expect new physics showers from signal yields under these cuts. Huge {charged CR} backgrounds can be removed by $X_0^{\rm cut}$ {(See section \ref{subsec:X0cut})}, and new physics showers can be distinguished from remaining CC/NC shower backgrounds by considering $\Delta X^{\rm cut}$. If we consider the view in Fig. \ref{EAS_view_schematic}, we estimate the maximum position as follows \cite{Mollerach:2017idb}:
\begin{eqnarray}
X_{\max} & \simeq & X_0 + X_{\rm MFP}^{\pi p} \ln \Bigl ( \frac{E_{\rm CR}}{N_{\pi^\pm}^{(0)} \cdot E_{\rm crit.}} \Bigr )
\end{eqnarray}
where $N_{\pi^\pm}^{(0)}$ is the number of charged pions produced in the first hard interactions indicated in Fig. \ref{Charged_Particle_multiplicity_primary_int} for each interactions, and $X_{\rm MFP}^{\pi p} = 120$ g/cm$^2$ is the mean-free-depth of pions in the atmosphere. $E_{\rm crit.}$ is the minimum energy of charged pions satisfying $\gamma_\pi c \tau_{\pi^\pm} < $ (layer size), in order to make the pions reinteract with nucleons in air molecules before they decay into muons and neutrinos. As a result, the difference $X_{\max}^{\rm CC/NC} - X_{\max}^{\rm NP}$ is given by
\begin{eqnarray}
X_{\max}^{\rm CC/NC} - X_{\max}^{\rm Sph,BH} & \simeq & {X_{\rm MFP}^{\pi p}} \ln \left ( \frac{N_{\pi^\pm (\text{Sph,BH})}^{(0)}}{N_{\pi^\pm (\text{CC/NC})}^{(0)}} \right ) \ \ \ 
\end{eqnarray}

Clearly, we notice that the value $\ln ( \frac{E_{\rm CR}}{N_{\pi^\pm}^{(0)} \cdot E_{\rm crit.}} )$ corresponds to the number of layers in Fig. \ref{EAS_view_schematic}.

%%%%%%%%%%%%%%%%%%%%%%%%%%%%% 
\subsection{Cutting the charged CR background events}
\label{subsec:X0cut}
%%%%%%%%%%%%%%%%%%%%%%%%%%%%% 

 In general, the charged CR background events from the collision of UHECR proton/heavy nuclei and nucleon in the atmosphere are dominant sources of shower events observed at a ground based experiment. However, imposing an aggressive cut at a large $X_0$ greatly reduces the background as observed in Ref.~\cite{Anchordoqui:2001cg}. 
 Taking the short mean-free-depth $\lambda_{{\rm QCD}} \sim 80$ g/cm$^2$ for the QCD event, we found that the cut $X_0^{\rm cut} \gsim 1200$ g/cm$^2$ removes almost all QCD backgrounds. The neutrino events, by new physics (black hole, sphaleron) or CC/NC, on the other hand, have longer mean-free-depth and the event distribution is uniform thus remains relatively large as clearly seen in Fig.~\ref{NevAfterX0cut}.
 
%%%%%%%%%%%%%%%%%%%%%%%%%%%%%%%%%%%%%%%%%%%
\begin{figure}[h]
\centering
\includegraphics[width=.46\textwidth]{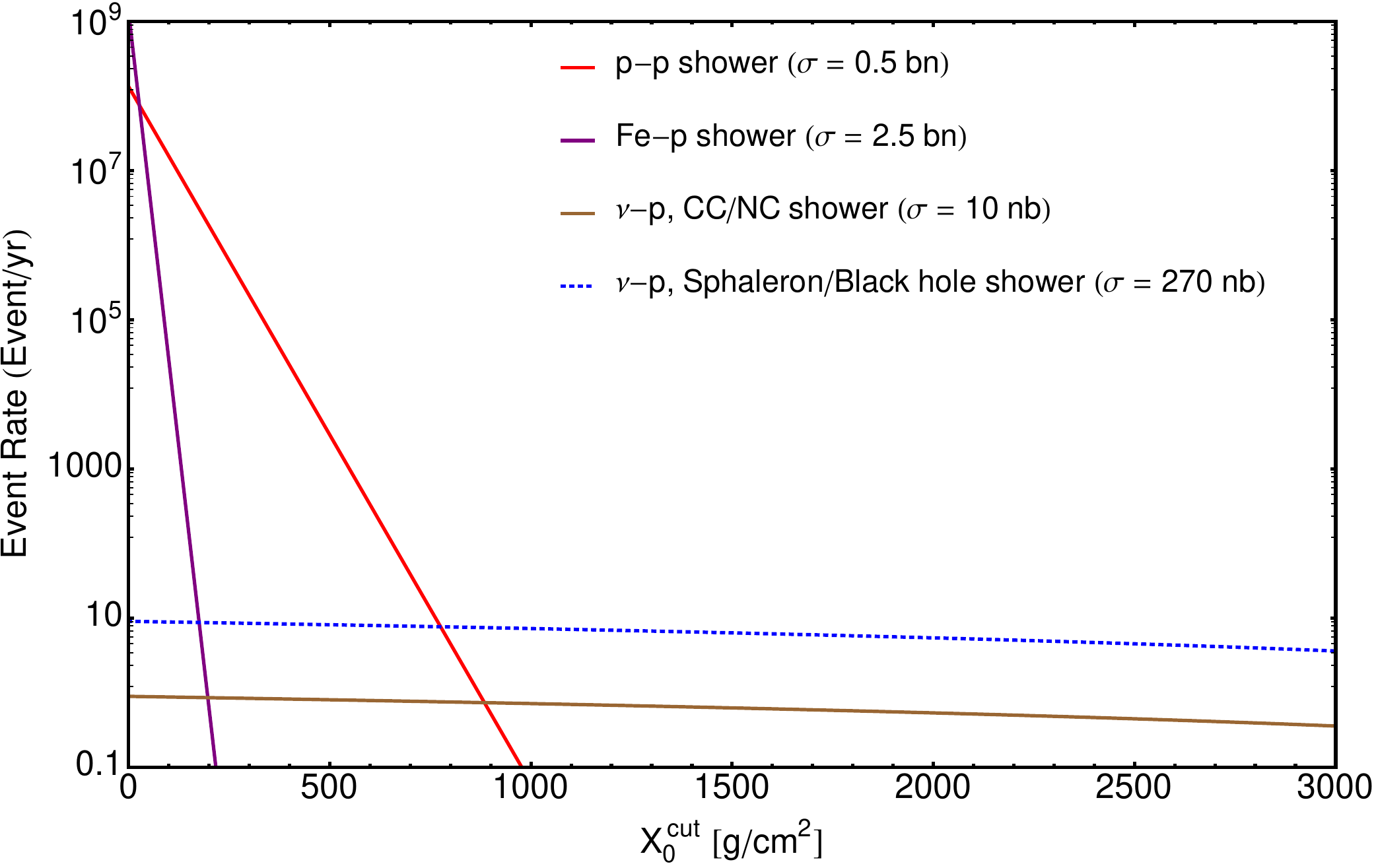}
\caption{{\bf Event rate on ground detector array after $X_0$ cut}. Expected number of events per year on Auger after injection point cut $X_0^{\rm cut}$ for primary particle energy $10^8 \ {\rm GeV} < E_{\rm CR} < 10^{11}$ GeV. New physics event is neutrino-nucleon collision induced by electroweak sphaleron (Eq. \ref{eq:sph} with $p=1.0$). Total event numbers are obtained for the maximum slant depth $X = 5000$ g/cm$^2$ which corresponds to $\theta_{\rm zenith} = 80^\circ$. We use (\ref{X0dist}) for $X_0$ distribution with $\sigma_{\rm int}$ as each cross sections indicated in the figure.} \label{NevAfterX0cut}
\end{figure}
%%%%%%%%%%%%%%%%%%%%%%%%%%%%%%%%%%%%%%%%%%%

%%%%%%%%%%%%%%%%%%%%%%%%%%%%% 
\subsection{The muon number in the UHECR events}\label{airShowersAugerMuonExcessDiscussion}
%%%%%%%%%%%%%%%%%%%%%%%%%%%%%

Finally, in this {\color{red} sub}section, we address the anomalous result reported by the Pierre Auger collaboration with their 9-years data~\cite{Aab:2014pza,Aab:2014dua} and see the potential account from the new physics effects. The 9-years (2004-2013) Pierre Auger Observatory data set contains many UHE cosmic ray air-shower events, including 29,722 `highly inclined events' with the  event selection criteria:
\begin{itemize}
\item $62^\circ < \theta_{\rm zenith} < 80^\circ$ with $\theta_{\rm zenith}^{\rm avg.} = 67^\circ$,
\item $E_{\mu^\pm} > 0.3$ GeV%, which is the Cherenkov threshold for muons in water,
\item $E_{\rm CR} \geq 5 \times 10^{18}$ eV% corresponding to $\sqrt{s} \geq 100$ TeV in the center-of-mass frame.
\end{itemize}
Among these 29,722 events, there are 174 events indicate the deficit of muons in the Monte Carlo simulations. These highly inclined shower events have more muons than the expected number of muons predicted by the hadronic models. Importantly, the muon number is detectable via SD/FD hybrid detection \cite{Aab:2014pza,Aab:2014dua}. Specifically, $R_\mu = N_\mu / N_{\mu ,19}$ parameter is used to define the normalized muon number in each extensive air shower event: the total number of muons in each event divided by the reference value of the muon number. We take the reference value $N_{\mu,19} = 2.68 \times 10^7$ at $\theta_{\rm zenith} = 67^\circ$ from Ref.~\cite{Aab:2014pza}, which is obtained from the MC simulation at $E_{\rm CR} = 10^{19}$ eV. We take these events seriously since the similar muon issues are also found in TA 7-years (2008-2015) data set \cite{Abbasi:2018fkz}. The deficit in interaction models mainly appear in highly inclined and highest energies air-showers \cite{Fomin:2016kul}.
To account the muon issue in air-shower events, several approaches have been proposed even though no thorough explanation has been given, including the revision of hadronic interaction models at high energy collisions in $\sqrt{s} = 110 - 170$ TeV range~\cite{Aab:2016hkv} as well as new physics contributions \cite{Farrar:2013sfa,AlvarezMuniz:2012dd}.
 
 We check if the new physics interactions due to sphaleron or black hole improves the situation. The results are shown in Fig.~\ref{MuonExcessComparison}. The new physics interactions indeed provide some enhancement 
compared with the QCD events (proton, iron with different modelings) at high energies, $E>10^{9.5}$ GeV.  However, when the muon number is averaged over the energy ranges, $R_\mu^{{\rm avg.}} (E_{{\rm CR}})$, the new physics interactions do not provide any significant enhancement except for the very deep injection from  $h_{\rm inj} \lsim 5.0$ km (or equivalently, $X_0 \gsim 1400$ g/cm$^2$), which is not very likely for the new physics. In conclusion, the new physics interactions -by sphaleron and black hole- do not seem to explain the muon deficit in the air showers generated by the hadronic interaction models. 

As we mention in Appendix \ref{subsec:X0cut}, the muon number measurement for highly-inclined air-showers still can be informative for new physics search after imposing the $X_0$ cut.

%%%%%%%%%%%%%%%%%%%%%%%%%%%%%%%%%%%%%%%%%%%
\begin{figure}[h]
\centering
\includegraphics[width=0.45\textwidth]{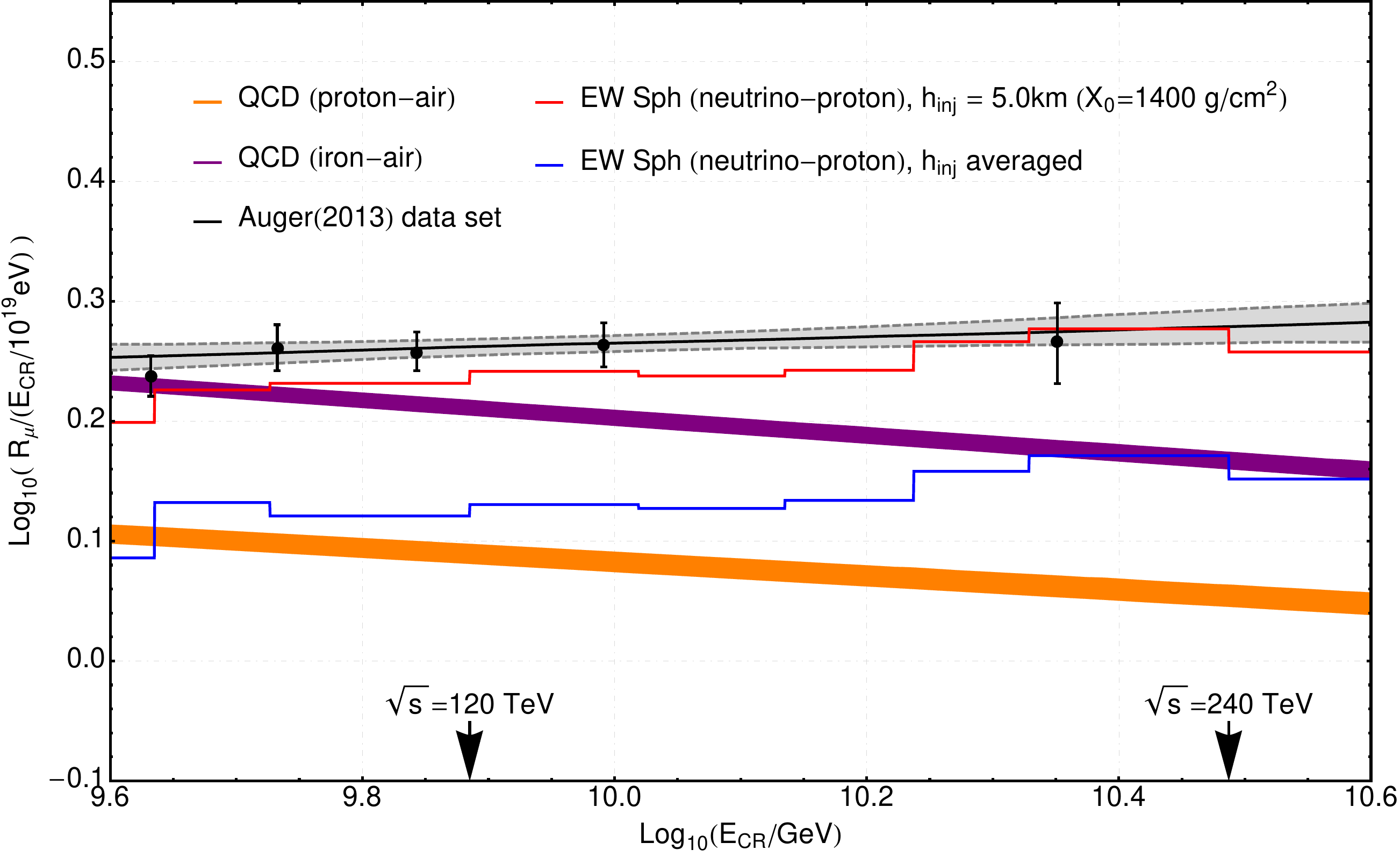}
\caption{The total number of muons in highly-inclined air-showers of 9-years Auger data \cite{Aab:2014pza} and the expected muon numbers for sphaleron-induced air-showers for a fixed initial injection height $h_{\rm inj} = 5.0$ km of extensive air showers (red solid line). The depth-averaged muon number is also shown (blue solid line). \label{MuonExcessComparison}}
\end{figure}
%%%%%%%%%%%%%%%%%%%%%%%%%%%%%%%%%%%%%%%%%%%

%%%%%%%%%%%%%%%%%%%%%%
\section{Conclusion \label{sec:conclusion}}
%%%%%%%%%%%%%%%%%%%%%%

Targeting potential new physics located at $\mathcal{O}(10)$ TeV, we studied generic search strategies at the air-shower detector array experiments such as Telescope Array Experiment and Pierre-Auger Observatory, where the collisions of { UHE neutrinos} and nucleons in the atmosphere are observed. Comparing the signal events with conventional { charged CR and CC/NC} background events, we found that our target events have larger multiplicities and thus have characteristic features in showering processes: { {\it i)} larger $X_0$ than charged CR background, {\it ii)} larger $N_{\rm Charged}$ than CC/NC neutrino shower at higher energies, {\it iii)} smaller $\Delta X$ than CC/NC background}, and also distinguishable numbers of electromagnetic, muonic and hadronic components. The features are highlighted in the schematic figure in Fig.~\ref{EAS_view_schematic}. Finally, we also studied potential implication of new physics interactions to the `muon deficit in models' seen in Auger and TA data. %Some details of our analysis are presented in the appendices.

%When the improved determination of the mass composition \cite{Aab:2017cgk} is achieved and the origin of UHECR is better clarified in the future \cite{Aab:2017tyv}, the {\color{red} UHE neutrino} air-shower events from the high-multiplicity processes can be better probed based on our analysis. 

We expect the future air-shower array experiments such as TA$\times4$ \cite{Sagawa:2016ysc}, FAST \cite{Fujii:2015dra}, GRAND \cite{Fang:2017mhl} POEMMA \cite{Olinto:2017xbi}, and AugerPrime \cite{Aab:2016vlz} will reveal the nature of physics at TeV and beyond.

\begin{acknowledgements}
This work was supported by the National Research Foundation of Korea (NRF) grants funded by the Korean government (MSIP) (2021R1A4A2001897) and (2019R1A2C1089334)
\end{acknowledgements}

\appendix

\section{Event rate in the air-shower detector array} \label{EvRateAirshowerCCNCNP}

\subsection{CC/NC neutrino-induced deep inclined air-shower event}

The CC/NC neutrino-induced, nearly horizontal deep air-shower event rates on the Pierre Auger, for each shower origin, are \cite{Gandhi:1998ri}
\begin{itemize}
\item NC shower for all 3 flavors $\nu_{l=e,\mu,\tau}$ and $\bar{\nu}_{l=e,\mu,\tau}$. ($\nu_l q \rightarrow \nu_l q^*$)
\begin{eqnarray}
\frac{dN_\nu^{NC}}{dt} & = & \rho_{\rm Air} \ N_A \sum_{l=e,\mu,\tau} \int_{E_{\min}}^\infty dE_{\rm sh} \frac{d\phi_{\nu_l} (E_\nu)}{d E_\nu} \nonumber \\
& & \times  \int_{y_{\min}}^{y_{\max}} dy \ \frac{d\sigma_{\nu_l}^{NC}(E_\nu, y)}{dy} \ \mathcal{A} (E_{\rm sh})
\end{eqnarray}
where $E_{\rm sh} = y E_\nu = E_\nu - E_\nu '$ is the total hadronic shower energy.

\item CC shower for $\nu_e$ and $\bar{\nu}_e$. ($\nu_e q \rightarrow e q'$)
\begin{eqnarray}
\frac{dN_{\nu_e}^{CC}}{dt} & = & \rho_{\rm Air} \ N_A \int_{E_{\min}}^\infty dE_{\rm sh} \ \frac{d\phi_{\nu_e} (E_\nu)}{d E_\nu} \nonumber \\
& & \times \int_0^1 dy \ \frac{d\sigma_{\nu_e}^{CC}(E_\nu, y)}{dy} \ \Theta (E_{\max} - E_\nu) \ \mathcal{A}(E_{\rm sh}) \nonumber \\
\end{eqnarray}
where $y E_\nu = E_\nu - E_e$ and $E_e$ is the hadronic shower and EM shower energy, respectively. The total shower energy is $E_{\rm sh} = E_\nu$.

\item CC shower for $\nu_\tau$ and $\bar{\nu}_\tau$ with hadronically decaying $\tau$ ($\nu_\tau q \rightarrow \tau q'$, and $\tau \rightarrow \nu_\tau q''\bar{q}'''$) 
\begin{eqnarray}
\frac{dN_{\nu_\tau}^{CC{\rm -had}}}{dt} & = & \rho_{\rm Air} \ N_A \int_{E_{\min}}^\infty dE_{\rm sh} \ \frac{d \phi_{\nu_\tau}(E_\nu)}{d E_\nu} \nonumber \\
& & \times \int_0^1 dy \ \frac{d\sigma_{\nu_\tau}^{CC}(E_\nu, y)}{dy} \int_0^1 dz \ \frac{dn(\tau \rightarrow {\rm had})}{dz} \nonumber \\
& & \times \Theta \Bigl ( E_\nu (y+ (1-y)(1-z) ) - E_{\min} \Bigr ) \nonumber \\
& & \times \Theta \Bigl ( E_{\max} - E_\nu (y+ (1-y)(1-z)) \Bigr ) \nonumber \\
& & \times {\rm Br} (\tau \rightarrow {\rm hadrons}) \times \mathcal{A} (E_{\rm sh})
\end{eqnarray}
where $y E_\nu = E_\nu - E_\tau$ is the deposited energy, and $z = E_\nu ' / E_\tau$ is the fraction of invisible (neutrino) energy from tau decays and total shower energy is the sum of hadronic energy of the broken nucleon $y E_\nu$ and hadronic energy from tau decays $(1-y)(1-z) E_\nu$.

\item CC shower for $\nu_\tau$ and $\bar{\nu}_\tau$ with electronically decaying $\tau$ ($\nu_\tau q \rightarrow \tau q'$, and $\tau \rightarrow \nu_\tau \bar{\nu}_e e$) 
\begin{eqnarray}
\frac{dN_{\nu_\tau}^{CC{\rm -em}}}{dt} & = & \rho_{\rm Air} \ N_A \int_{E_{\min}}^\infty dE_{\rm sh} \ \frac{d \phi_{\nu_\tau}(E_\nu)}{dE_\nu} \nonumber \\
& & \times \int_0^1 dy \frac{d\sigma_{\nu_\tau}^{CC}(E_\nu,y)}{dy} \int_0^1 dz \ \frac{dn (\tau \rightarrow \nu_\tau \bar{\nu}_e e)}{dz} \nonumber \\
& & \times \Theta \Bigl ( E_\nu (y+(1-y)z) - E_{\min} \Bigr ) \nonumber \\
& & \times \Theta \Bigl ( E_{\max} - E_\nu (y+(1-y)z) \Bigr ) \nonumber \\
& & \times {\rm Br} (\tau \rightarrow \nu_\tau \bar{\nu}_e e) \times \mathcal{A} (E_{\rm sh})
\end{eqnarray}
where $y E_\nu = E_\nu - E_\tau$ is the deposited energy again, and $z = E_e / E_\tau$ is the fraction of EM shower energy from tau decays and total shower energy is the sum of hadronic energy of broken nucleon $y E_\nu$ and tau EM shower energy $(1-y)z E_\nu$.

\item CC shower for $\nu_\mu$ and $\bar{\nu}_\mu$. ($\nu_\mu q \rightarrow \mu q'$) 
\begin{eqnarray}
\frac{dN_{\nu_\mu}^{CC{\rm -had}}}{dt} & = & \rho_{\rm Air} \ N_A \int_{E_{\min}}^\infty dE_{\rm sh} \ \frac{d\phi_{\nu_\mu}(E_\nu)}{dE_\nu} \nonumber \\
& & \times \int_{y_{\min}}^{y_{\max}} dy \ \frac{d\sigma_{\nu_\mu}^{CC}(E_\nu, y)}{dy} \ \mathcal{A} (E_{\rm sh}) \ \ \ \
\end{eqnarray}

where $E_{\rm sh} = y E_\nu = E_\nu - E_\mu$ is the total shower energy.

\item CC shower for $\nu_\tau$ and $\bar{\nu}_\tau$ with muonically decaying $\tau$ ($\nu_\tau q \rightarrow \tau q'$, and $\tau \rightarrow \nu_\tau \bar{\nu}_\mu \mu$) 
\begin{eqnarray}
\frac{dN_{\nu_\tau}^{CC{\rm -had}}}{dt} & = & \rho_{\rm Air} \ N_A \int_{E_{\min}}^\infty dE_{\rm sh} \ \frac{d\phi_{\nu_\tau}(E_\nu)}{dE_\nu} \nonumber \\
& & \times {\rm Br}(\tau \rightarrow \nu_\tau \bar{\nu}_\mu \mu) \nonumber \\
& & \times \int_{y_{\min}}^{y_{\max}} dy \ \frac{d\sigma_{\nu_\tau}^{CC}(E_\nu, y)}{dy} \ \mathcal{A}(E_{\rm sh}) \ \ \ \ 
\end{eqnarray}
where $E_{\rm sh} = y E_\nu = E_\nu - E_\tau$ is the total shower energy, again.
\end{itemize}

For the detailed evaluation, we need several quantities defining the detector size, the strength of each interaction and the flux of neutrinos from various sources such as
\begin{itemize}

\item $\mathcal{A} (E_{\rm sh})$ is the energy-dependent effective array acceptance{, obtained from the effective area \cite{Aab:2019dav, Aab:2019ogu} for UHE neutrino search with CC $\nu_e$ event and rescaled by cross sections for other CC/NC and high multiplicity processes.} Basically, the Pierre Auger detector array is sensitive above $\mathcal{O}(100)$ PeV, in which GZK neutrinos are dominant. 

\item $d\sigma_{\nu_l}^{CC,NC}/dy$ is the differential CC and NC neutrino ($\nu_l$)-nucleon ($N$) cross section in the SM \cite{Gandhi:1998ri} respectively, and the total cross sections are
\begin{eqnarray}
\sigma_{\nu_l}^{CC} (E_\nu) & = & \frac{2 G_f^2 M_N E_\nu}{\pi} \int_0^1 dy \int_0^1 dx \Bigl ( \frac{M_W^2}{Q^2 + M_W^2} \Bigr )^2\nonumber \\ 
& & \times \sum_q \Bigl [ x f_q (x, Q^2) + x f_{\bar{q}} (x, Q^2) (1-y)^2 \Bigr ] \ , \nonumber \\
\end{eqnarray}
\begin{eqnarray}
\sigma_{\nu_l}^{NC} (E_\nu) & = & \frac{G_F^2 M_N E_\nu}{2\pi} \int_0^1 dy \int_0^1 dx \ \Bigl ( \frac{M_Z^2}{Q^2 + M_Z^2} \Bigr )^2 \nonumber \\
& & \times \sum_q \Bigl [ xf_{q^0}(x,Q^2) + xf_{\bar{q}^0}(x,Q^2) (1-y)^2 \Bigr ] \ , \nonumber \\
&& 
\end{eqnarray}
where $x$ is the parton fraction in the nucleon, and $y$ is the fraction of deposited energy. $f_q (x, Q^2)$ and $f_{\bar{q}} (x, Q^2)$ are also defined in \cite{Gandhi:1998ri}.

\item The energy spectrums $dn/dz$ in electronically \cite{Scheck:1977yg} and hadronically \cite{Davier:1992nw} decaying $\tau$ and the branching ratio in the $\tau$ decay.

\item There are neutrinos due to the interaction between high-energy cosmic rays and Earth atmosphere nuclei \cite{Aartsen:2013vca} and the neutrinos of the astrophysical origin, such as highly-accelerated hadrons in supernovae remnants, active galactic nuclei (AGN), gamma-ray bursts (GRBs), and shocks in star formation regions of galaxies \cite{Aartsen:2015knd}.

\item Above $\mathcal{O}(100)$ PeV, the UHE neutrinos are produced by the interaction between the UHE cosmic rays (mainly protons, and small fractions of other heavy nuclei) above $\mathcal{O}(10^9)$ GeV and the CMB photons \cite{Greisen:1966jv,Zatsepin:1966jv}, by the following channels: $p + \gamma_{\rm bkg.} \rightarrow \pi^+ + n, \ \pi^+ \rightarrow \mu^+ + \nu_\mu$ and $p + \gamma_{\rm bkg.} \rightarrow \pi^0 + p, \ \pi^0 \rightarrow \gamma + \gamma$. This generation of pions is also denoted as the photo-pion production. The cross section for pion production has a resonance peak at the $\Lambda^+ (1232)$ resonance \cite{Andersen:2011dz}. These UHE neutrinos from the GZK mechanism are expected from the observations of the UHE cosmic rays by the air-shower detector arrays \cite{Valino:2015zdi,Tinyakov:2018hfg} and the observations of diffuse photons by the gamma-ray telescopes \cite{Ahlers:2010fw}, although there is no direct observation of the GZK neutrinos by the neutrino telescopes yet \cite{Aartsen:2015zva,Ishihara:2016pty}.
\end{itemize}

The UHE neutrinos can produce nearly horizontal and deep air-showers, which correspond to $X \sim 13,000$ g/cm${}^2$ \cite{Abreu:2013zbq}.
For typical flux values and CC/NC interactions, we expect $\sim (0.9-2.9)$ events/yr with typical choices of the acceptance values and the GZK neutrino flux models, although no neutrino-induced event candidates have been found yet \cite{Parente:1996bv}, which provides the bound on the GZK neutrino flux.

\subsection{{Sphaleron and black hole} air-shower event}

The event rate in the ground air-shower detector array is given by
\begin{eqnarray}
\frac{dN}{dt} & = & N_A \ \rho_{{\rm air}} \int_{E_{{\rm th}}}^{E_{\max}} dE_{{\rm sh}} \int_0^1 dy \nonumber \\
& & \times \frac{d \phi_{\nu_l} (E_\nu)}{d E_\nu} \ \frac{d \sigma_{\nu_l X} (E_\nu, y)}{dy} \ \mathcal{A} (E_{{\rm sh}}) \ .
\end{eqnarray}
where $\mathcal{A}(E_{{\rm sh}})$ is the air-shower energy-dependent acceptance of the entire detector array. For Auger, we adopt the values  obtained from the effective area \cite{Aab:2019dav, Aab:2019ogu} for CC $\nu_e$ UHE neutrino with rescaling by Sphaleron/black hole cross section.

\section{Mean-Free-Path weighted muon number} 
\label{MFPavgMuonNumber}

The mean-free-path averaged muon number, which is given by
\begin{eqnarray}
R_\mu^{{\rm avg.}} (E_{{\rm CR}}) & = & \int_0^{X_{\rm obs.}} dX_0 \ P \Bigl ( X_0, \ \sigma_{\rm int} (E_\nu) \Bigr ) \ R_\mu (X_0, \ E_\nu) \nonumber \\
\end{eqnarray}
where
\begin{eqnarray}
P(X_0, \sigma_{\rm int} (E_\nu) ) & = & \frac{1}{X_{\rm MFP} (\sigma_{\rm int}(E_\nu) )} \nonumber \\
& & \times \exp \Bigl (- X_0/ X_{\rm MFP} (\sigma_{\rm int}(E_\nu)  ) \Bigr ) \ \ \ \ \label{X0dist}
\end{eqnarray}
and $X_{\rm MFP} (\sigma_{\rm int}) = A_{\rm atm} \cdot N_A^{-1} \cdot \sigma_{\rm int}^{-1} (E_\nu)$ is the mean-free-depth. $A_{\rm atm}$ is the atomic mass of the atmosphere. Because the cross section for new physics $\sigma_{\rm int}$ is small enough for our parameter choices, $P(X_0, \sigma_{\rm int})$ is almost constant and the air-showers can occur everywhere with almost uniform distribution.

\bibliographystyle{JHEP}
\bibliography{refs}

\end{document}